\documentclass[10pt,twoside,twocolumn,english,superscriptaddress, aps, showkey]{revtex4-1}
\usepackage[T1]{fontenc}
\usepackage[latin9]{inputenc}
\usepackage{babel}
\usepackage{amsmath}
\usepackage{graphicx}
\usepackage[unicode=true]
 {hyperref}

\makeatletter

\providecommand{\tabularnewline}{\\}

\date{\today}
\usepackage{makecell}

\@ifundefined{showcaptionsetup}{}{%
 \PassOptionsToPackage{caption=false}{subfig}}
\usepackage{subfig}
\makeatother

\begin{document}
\author{Yingchao Lu}
\email{yclu@lanl.gov}
\affiliation{Theoretical Division, Los Alamos National Laboratory, Los Alamos, New Mexico 87545, USA}
\affiliation{Department of Physics and Astronomy, Rice University, Houston, Texas 77005, USA}
\author{Shengtai Li}
\email{sli@lanl.gov}
\author{Hui Li}
\email{hli@lanl.gov}
\affiliation{Theoretical Division, Los Alamos National Laboratory, Los Alamos, New Mexico 87545, USA}
\author{Kirk A. Flippo}
\email{kflippo@lanl.gov}
\author{Dan Barnak}
\affiliation{Laboratory for Laser Energetics,  Rochester, New York 14623, USA}
\author{Andrew Birkel}
\author{Brandon Lahmann}
\author{Chikang Li}
\affiliation{Plasma Science and Fusion Center, Massachusetts Institute of Technology, Cambridge, Massachusetts 02139, USA}
\author{Alexander M. Rasmus}
\author{Kwyntero Kelso}
\affiliation{Physics Division, Los Alamos National Laboratory, Los Alamos, New Mexico 87545, USA}
\author{Alex Zylstra}
\affiliation{Lawrence Livermore National Laboratory, Livermore, California 94550, USA}
\author{Edison Liang}
\affiliation{Department of Physics and Astronomy, Rice University, Houston, Texas 77005, USA}
\author{Petros Tzeferacos}
\author{Don Lamb}
\affiliation{Department of Astronomy and Astrophysics, University of Chicago, Chicago, Illinois 60637, USA}
\title{Modeling hydrodynamics, magnetic fields and synthetic radiographs
for high-energy-density plasma flows in shock-shear targets}
\begin{abstract}
Three-dimensional FLASH radiation-magnetohydrodynamics (radiation-MHD)
modeling is carried out to study the hydrodynamics and magnetic fields
in the shock-shear derived platform. Simulations indicate that fields
of tens of Tesla can be generated via Biermann battery effect due
to vortices and mix in the counter-propagating shock-induced shear
layer. Synthetic proton radiography simulations using MPRAD and synthetic
X-ray image simulations using SPECT3D are carried out to predict the
observable features in the diagnostics. Quantifying the effects of
magnetic fields in inertial confinement fusion (ICF) and high-energy-density
(HED) plasmas represents frontier research that has far-reaching implications
in basic and applied sciences.
\end{abstract}

\maketitle

\section{Introduction}

When an inertial confinement fusion (ICF) capsule implodes, the material
turns into dense plasmas and recent simulations have shown that such
plasmas tend to be unstable and turbulence can develop\citep{ICF_HS_PhysRevE.89.053106}.
Even though it is debated whether turbulence is damped by the viscosity
in the hot spot, the shocked interfaces as well as the interface between
the shell and the hot spot can have very different dynamics and can
indeed be unstable\citep{ICFhs_Clark2015,ICFhs_Haines2014,ICFhs_Haines2016,ICFhs_Haxhimali2015,ICFhs_Vold2015,ICFhs_Vold2016}.
It is believed that turbulence and the associated mixing process can
be crucial for understanding ICF.

The Biermann battery effect\citep{Biermann1950} is known to generate
seed magnetic fields in laser driven plasma flows and has been studied
extensively in high-energy-density (HED) laser-driven experiments\citep{Prad_Li2006,PRAD_BB_Cecchetti2009,PRAD_BB_Gao2019,PRAD_BB_Li2007,PRAD_BB_Li2009,PRAD_BB_Li2013,PRAD_BB_Petrasso2009,BB_Gregori2012},
but the strength and importance of these fields in the close to or
higher than solid density plasmas such as an ICF implosion are not
well known. Three-dimensional extended-magnetohydrodynamic (extended-MHD)
simulations of the stagnation phase of ICF including Biermann battery
term\citep{Biermann1950}, Nernst term\citep{NERNST_Nishiguchi1984}
and anisotropic heat conduction in the magnetic field, indicate that
self-generated magnetic fields can reach over $10^{4}$ Tesla and
can affect the electron heat flow\citep{GORGON_Walsh2017}. The simulations
with pre-magnetization for ICF implosions show the significance of
Lorentz force and $\mathrm{\alpha}$-particle trapping\citep{GORGON_Walsh2019}.
In low density laser driven plasmas, the magnetic field can be amplified
by turbulence and measured using temporal diagnostics by B-dot probe\citep{TMD_Meinecke2014}
and spatial diagnostics by proton radiography\citep{TMD_Tzeferacos2018}.
The magnetic frequency spectrum in supersonic plasma turbulence has
been measured in a recent experiment\citep{TMD_White2019} on the
Vulcan laser. However, in those experiments\citep{TMD_Meinecke2014,TMD_Tzeferacos2018,TMD_White2019}
the magnetic field is not high enough to change the dynamics of the
hydrodynamical flow.

In this work, we use the shock-shear platform\citep{SS_Welser-Sherrill2013,SS_Capelli2016}
developed at Los Alamos National Laboratory (LANL) to quantify the
dynamics of magnetic fields in HED plasmas with instabilities and
turbulence. The shock compression can achieve a regime where the density
is around $1\mathrm{g/cc}$. The targets with large density can diffuse
the proton beam and affect the interpretation of the proton image\citep{MPRAD_Lu2019},
but the simulations for the synthetic proton image including the stopping
power and Coulomb scattering show that the deflection of proton beam
by magnetic fields is still detectable. Further improvements are still
needed to make the fields high enough to change the dynamics of the
small-scale evolution of vortices like those in a turbulent cascade,
and affect our understanding of turbulence.

The shock-shear platform\citep{SS_Welser-Sherrill2013,SS_Capelli2016},
as a platform to isolatedly study the shear-induced instabilities
and turbulence production under HED conditions, i.e. pressure larger
than $1\mathrm{Mbar}$, has been used to investigate the turbulent
mixing\citep{SS_Flippo2016b,SS_Flippo2016a} at material interfaces
when subject to multiple shocks and reshocks or high-speed shear\citep{SS_Welser-Sherrill2013,SS_Merritt2017}.
The experiments\citep{SS_Doss2015,SS_Merritt2015,SS_Flippo2018,SS_Doss2016b,SS_Doss2016a}
using the shock-shear platform has been carried out on the OMEGA Laser
Facility and National Ignition Facility (NIF). These experiments provide
quantitative measurements to assist in validation efforts\citep{SS_Doss2013a,SS_Doss2013b,SS_Wang2015}
for mix models, such as Besnard-Harlow-Rauenzahn (BHR) model\citep{BHR_Banerjee2010,BHR_Haines2013}.
The experimental data and the validation efforts constrain models
relevant to integrated HED experiments such as ICF or astrophysical
problems. In the shock-shear targets, the Biermann Battery ($\nabla n_{e}\times\nabla T_{e}$)
term\citep{Biermann1950} can generate and sustain strong magnetic
fields in the vortices due to the misalignment of the density gradient
and temperature gradient caused by electron heat conduction. However,
the magnetic fields in the shock-shear targets have not been quantified
in simulations or experiments.

In this work, we use the radiation-MHD code FLASH\citep{FLASH_Fryxell2000,FLASH_Dubey2009}
to model the evolution of the shock-shear system on OMEGA\citep{OMEGA_Boehly1997}.
The experiment simulated in this paper uses 8 beams each with 500J
energy laser ablation in 1$\mathrm{ns}$ on each side to drive strong
adjacent contour-propagating shocks. Kelvin-Helmholtz instability
laterally spreads across a thin layer of magnesium, copper or plastic
placed at the interface. The layer is cut with slots to seed the initial
density perturbation, which can generate vortices during the evolution
of the shock and shear. The temperature of the materials reaches tens
of electron-volts, and simulations predict the Mach number of the
post-shock flows in the experiment is around 2 on each side of the
shear layer. The magnetic field is generated by the Biermann battery
term\citep{Biermann1950} and dissipated by the resistive term. The
X-ray image\citep{XRFC_Benedetti2012,XRFC_Bradley1992,XRFC_Bradley1995}
and the proton radiography\citep{Prad_Li2006} are predicted and will
be compared to the experimental data in a later paper.

This paper is organized as follows. Sec \ref{sec:methods-and-config}
describes simulation methods and the configuration of the target system.
In Sec \ref{sec:Simulation-results}, we show the results for hydrodynamics
and MHD evolution from FLASH, the synthetic X-ray image using SPECT3D
and the synthetic proton radiography using MPRAD. The conclusions
and discussions is given in Sec \ref{sec:Conclusions-and-discussions}.

\section{Simulation methods and configuration\label{sec:methods-and-config}}

The FLASH code\citep{FLASH_Fryxell2000,FLASH_Dubey2009}\footnote{FLASH4 is available at \href{https://flash.uchicago.edu/}{https://flash.uchicago.edu/}}
is used to carry out the detailed physics simulations of our laser
experiments to study the dynamics of the shock-shear system. FLASH
is a publicly available, multi-physics, highly scalable parallel,
finite-volume Eulerian code and framework whose capabilities include:
adaptive mesh refinement (AMR), multiple hydrodynamic and MHD solvers\citep{Solver_Roe1981,Solver_Li2005,Solver_Miyoshi2005,Solver_Toro2009},
implicit solvers for diffusion using the HYPRE library\citep{HYPRE_Falgout2002}
and laser energy deposition. FLASH is capable of using multi-temperature
equation of states and multi-group opacities. To simulate laser-driven
High-Energy-Density-Physics (HEDP) experiments, a 3T treatment, i.e.
$T_{\mathrm{rad}}\ne T_{\mathrm{ele}}\ne T_{\mathrm{ion}}$, is usually
adopted. The equations which FLASH solves to describe the evolution
of the 3T magnetized plasma are

{\footnotesize{}
\begin{align}
\frac{\partial\rho}{\partial t}+\nabla\cdot(\rho\boldsymbol{v}) & =0\\
\frac{\partial\rho\boldsymbol{v}}{\partial t}+\nabla(\rho\boldsymbol{v}\boldsymbol{v}-\frac{1}{4\pi}\boldsymbol{B}\boldsymbol{B})+\nabla P_{\mathrm{tot}} & =0\\
\frac{\partial\rho E_{\mathrm{tot}}}{\partial t}+\nabla\cdot(\boldsymbol{v}(\rho E_{\mathrm{tot}}+P_{\mathrm{tot}})-\frac{1}{4\pi}\boldsymbol{B}(\boldsymbol{v}\cdot\boldsymbol{B}))\nonumber \\
-\frac{1}{4\pi}\nabla\cdot(\boldsymbol{B}\times(\eta\nabla\times\boldsymbol{B}))-\frac{1}{4\pi}\nabla\cdot(\boldsymbol{B}\times\frac{c}{e}\frac{\nabla P_{e}}{n_{e}}) & =-\nabla\cdot\boldsymbol{q}+S\\
\frac{\partial\boldsymbol{B}}{\partial t}+\nabla\cdot(\boldsymbol{vB}-\boldsymbol{Bv}) & =\nonumber \\
-\nabla\times(\eta_{B}\nabla\times\boldsymbol{B})+\frac{c}{e}\nabla\times\frac{\nabla P_{e}}{n_{e}}\label{eq:induction-equation}
\end{align}
}where the total pressure is given by $P_{\mathrm{tot}}=P_{\mathrm{ion}}+P_{\mathrm{ele}}+P_{\mathrm{rad}}+\frac{1}{8\pi}B^{2}$,
and the total specific energy $E_{\mathrm{tot}}=e_{\mathrm{ion}}+e_{\mathrm{ele}}+e_{\mathrm{rad}}+\frac{1}{8\pi}B^{2}+\frac{1}{2}\boldsymbol{v}\cdot\boldsymbol{v}$.
The total heat flux $\boldsymbol{q}$ is the summation of electron
heat flux $\boldsymbol{q}_{e}=-\kappa\nabla T_{\mathrm{ele}}$ and
radiation flux $\boldsymbol{q}_{r}$, where $\kappa$ is the Spitzer
electron heat conductivity\citep{Spitzer1963,Braginskii1965}. The
flux-limit used for electron thermal conduction is set to be 6\% of
the free streaming flux $q_{FS}=n_{e}k_{B}T_{e}\sqrt{\frac{k_{B}T_{e}}{m_{e}}}$.
The first term on the R.H.S of Eq(\ref{eq:induction-equation}) contains
the Spitzer magnetic resistivity $\eta_{B}$\citep{Spitzer1963,Braginskii1965}.
The second term on the R.H.S of Eq(\ref{eq:induction-equation}) is
the Biermann Battery term, which generates the magnetic field even
if there is no seed magnetic field initially. The plasma has zero
initial magnetic field in the simulations. Because plasma beta $\beta$
is much larger than unity, the Hall term is neglectable and not included
in the simulations. The Biermann battery term is turned off in the
cells adjacent to the shock detected numerically\citep{SHOCKDETECTION_Balsara1999}.
The magnetic fields generation near the shock is not calculated because
of the convergence problem\citep{FLASH_BB_Graziani2015} for calculating
Biermann battery term on the Eulerian grid. The convergence problem
might be resolved on a Lagrangian grid. On the other hand, the shock
in this work is highly collisional and with small thickness compared
to the spatial resolution of proton radiography, thus the scale of
the magnetic field near the shock is too small to be detectable. The
energy equations for the three components are

{\footnotesize{}
\begin{align}
\frac{\partial}{\partial t}(\rho e_{\mathrm{ion}})+\nabla\cdot(\rho e_{\mathrm{ion}}\boldsymbol{v})+P_{\mathrm{ion}}\nabla\cdot\boldsymbol{v} & =\rho\frac{c_{v,\mathrm{ele}}}{\tau_{ei}}(T_{\mathrm{ele}}-T_{\mathrm{ion}})\begin{aligned}\end{aligned}
\label{eq:eion}\\
\frac{\partial}{\partial t}(\rho e_{\mathrm{ele}})+\nabla\cdot(\rho e_{\mathrm{ele}}\boldsymbol{v})+P_{\mathrm{ele}}\nabla\cdot\boldsymbol{v} & =\rho\frac{c_{v,\mathrm{ele}}}{\tau_{ei}}(T_{\mathrm{ion}}-T_{\mathrm{ele}})\nonumber \\
-\nabla\cdot\boldsymbol{q}_{\mathrm{ele}}+Q_{\mathrm{abs}}-Q_{\mathrm{emis}}+Q_{\mathrm{las}}+Q_{\mathrm{ohm}} & \qquad\label{eq:eele}\\
\frac{\partial}{\partial t}(\rho e_{\mathrm{rad}})+\nabla\cdot(\rho e_{\mathrm{rad}}\boldsymbol{v})+P_{\mathrm{rad}}\nabla\cdot\boldsymbol{v} & =\nabla\cdot\boldsymbol{q}_{\mathrm{rad}}-Q_{\mathrm{abs}}+Q_{\mathrm{emis}}\label{eq:erad}
\end{align}
}where $c_{v,\mathrm{ele}}$ is the electron specific heat, $\tau_{ei}$
the ion-electron Coulomb collision time. The $Q_{\mathrm{abs}}$ (absorption)
and $Q_{\mathrm{emis}}$ (emission) describes the energy transfer
between the electron and the radiation, which is modeled using the
multi-group flux-limited radiation diffusion. The laser absorption
term $Q_{\mathrm{las}}$ is computed using ray-tracing in the geometric
optics approximation via the inverse-Bremsstrahlung process. $Q_{\mathrm{ohm}}$
is the rate of electron energy increase due to Ohmic heating. The
auxiliary equations Eq(\ref{eq:eion})-(\ref{eq:erad}) are advanced
in time such that the distribution of energy change due to the work
and the total shock-heating is based on the pressure ratio of the
components, which is a method implemented in FLASH inspired by the
radiation-hydrodynamics code RAGE\citep{RAGE_Gittings2008,RAGE_Haines2017}.
We use the equation of state and opacity table from PROPACEOS\citep{HELIOS_MacFarlane2006}\footnote{Prism Computational Sciences PrOpacEOS Overview \href{http://www.prism-cs.com/Software/Propaceos/overview.html}{http://www.prism-cs.com/Software/Propaceos/overview.html}}
for modeling all the material properties in our target system.

\begin{figure*}
\subfloat[The far-view of the target system, including the shock tube, the gold
cone for shielding and the plastic end cap. The foam and the layer
are not shown.]{\includegraphics[scale=0.5]{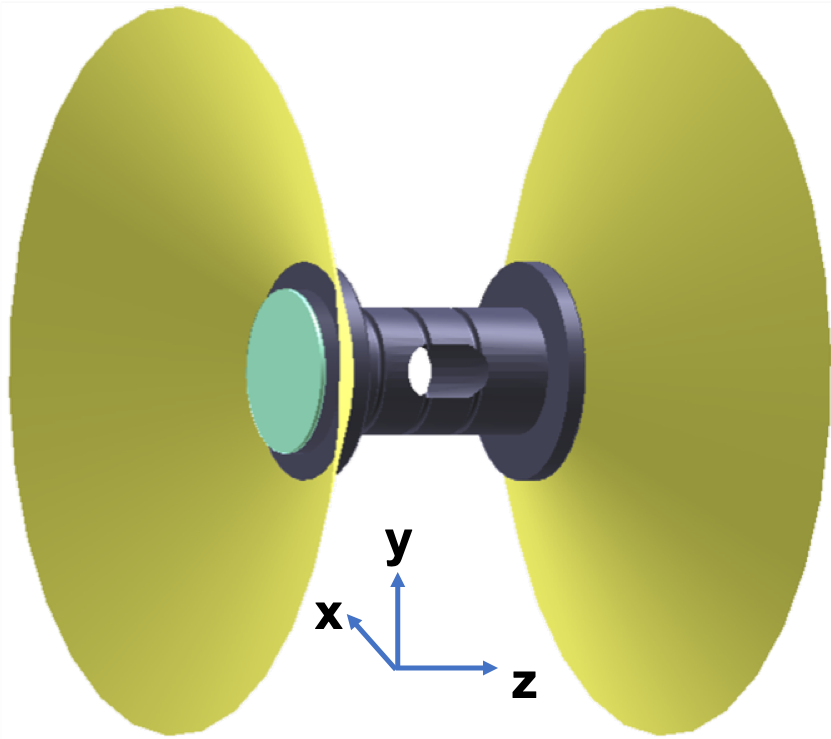}

}\subfloat[The target with a pepper-pot screen (PPS) for a narrow view proton
radiography. The screen has five large holes with $200\mathrm{\mu m}$
diameter and four small holes. The screen is at $x=-1.3\mathrm{mm}$
plane, attached to the edge of the gold cone. \label{fig:setup-pps}]{\includegraphics[scale=0.5]{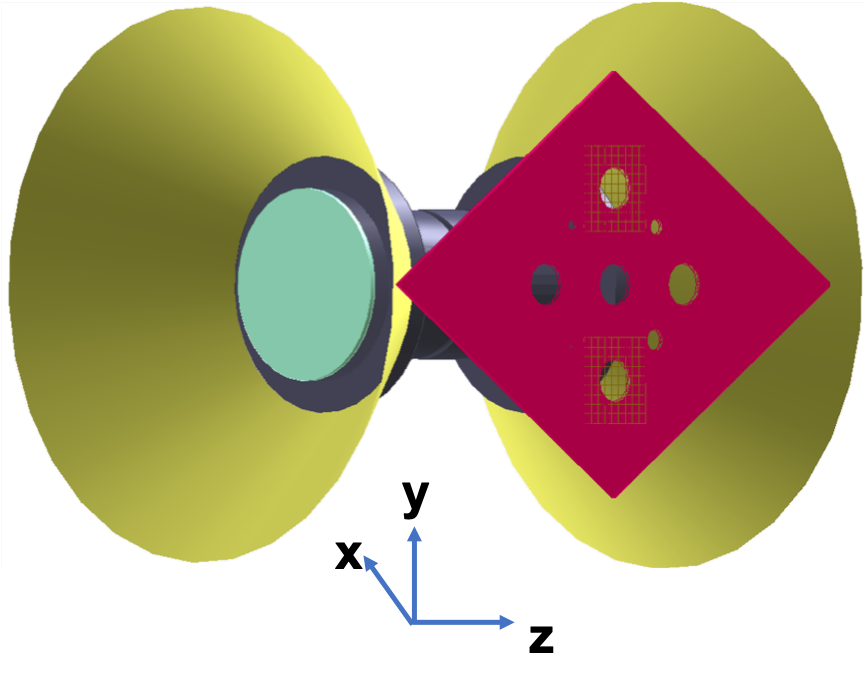}

}

\subfloat[The dimension of the shock tube, the window and the end cap. The beryllium
shock tube has a oval-shape window in the middle. The end cap is plastic.
The foam and the layer is not in this figure. The inner radius of
the tube is $250\mathrm{\mu m}$, and the outer radius of the tube
is $350\mathrm{\mu m}$. \label{fig:setup-tube}]{\includegraphics[scale=0.5]{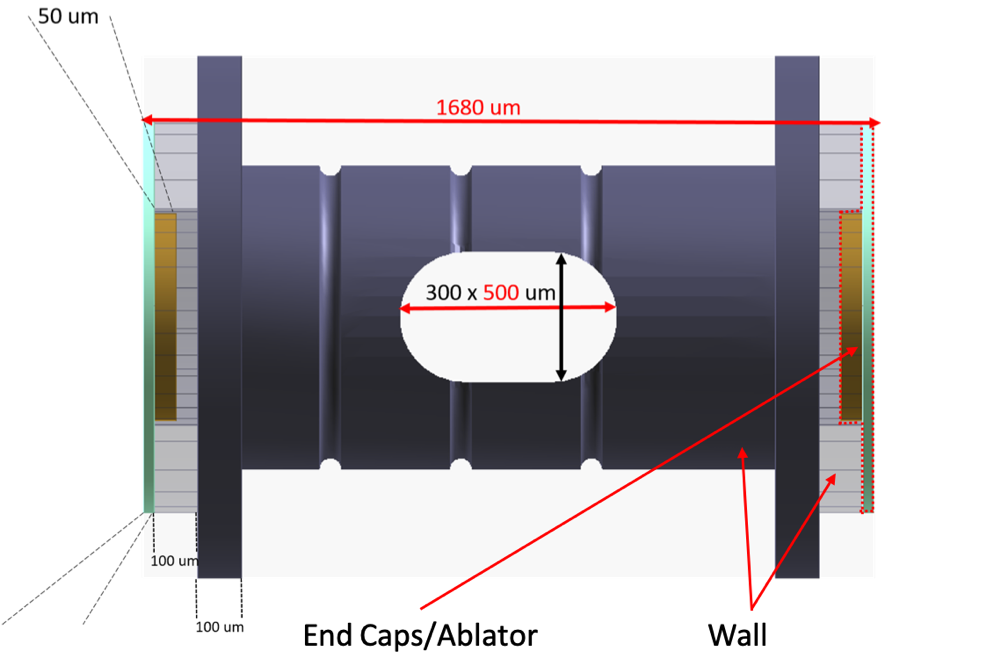}

}\subfloat[Same as (b) but the shock tube is plastic and thicker. The inner radius
of the tube is $250\mathrm{\mu m}$, and the outer radius of the tube
is $400\mathrm{\mu m}$. \label{fig:setup-tube2}]{\includegraphics[scale=0.5]{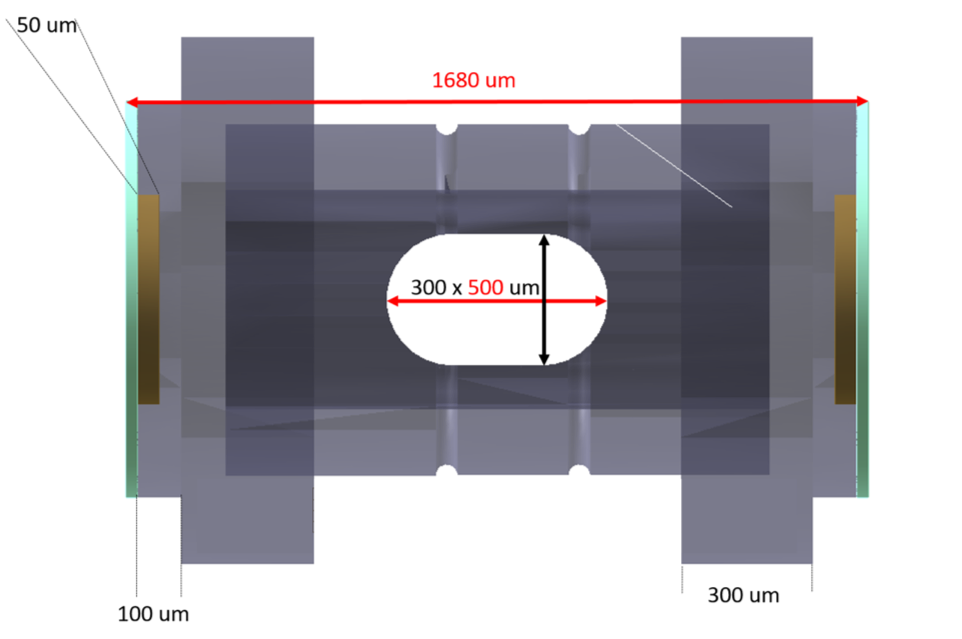}

}

\subfloat[The magnetism layer with 45 degree slanted slots. The wavelength of
the slots is $150\mathrm{\mu m}$. \label{fig:setup-layer1}]{\includegraphics[scale=0.35]{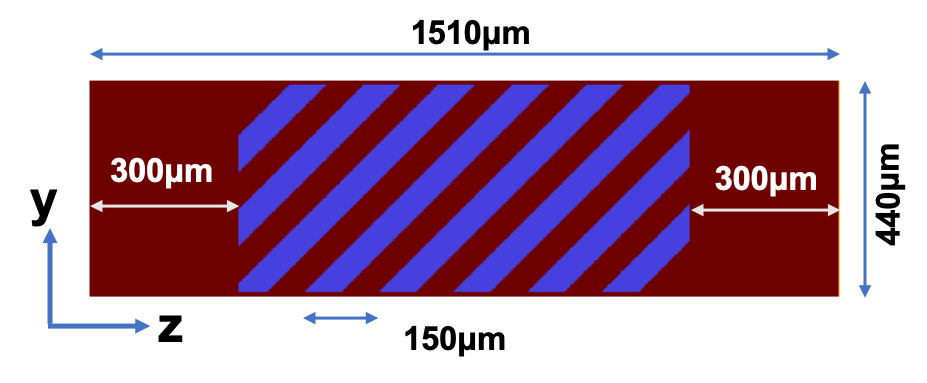}

}\subfloat[The plastic or copper layer with straight slots. The wavelength of
the slots is $150\mathrm{\mu m}$. \label{fig:setup-layer2}]{\includegraphics[scale=0.35]{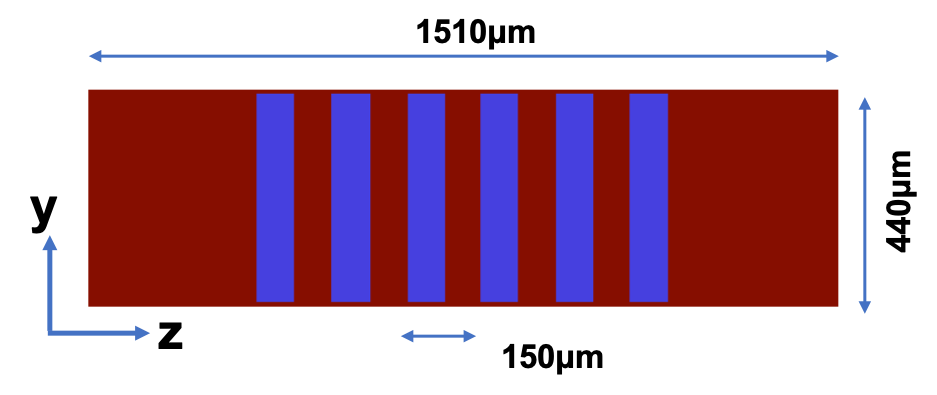}

}\subfloat[A layer divides the low density foam into two half-cylinders to collimate
the shock flow. The gold plugs hold back the shock at one end of each
half-cylinder of foam.\label{fig:setup-plug}]{\includegraphics[scale=0.35]{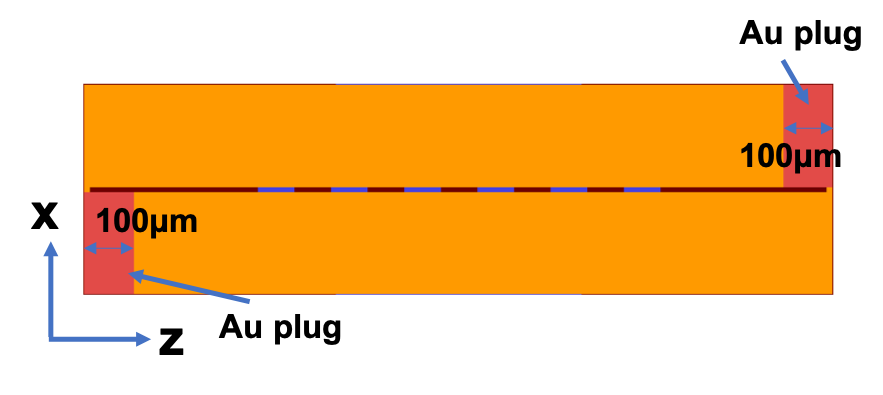}

}

\caption{The experiment setup. The shapes and dimensions of different parts
of the target is used to initialize the FLASH simulations. \label{fig:setup}}
\end{figure*}

We initialize the FLASH simulation using the geometry and parameters
of targets used for OMEGA experiments. The target system is composed
of the shock tube, the gold cone for minimizing stray laser light,
the foam filling the shock tube and a plastic cap covering the end
of the tube, as shown in Fig. \ref{fig:setup}. As shown in Fig. \ref{fig:setup-tube},
a window is opened in the middle of the tube and along the path of
the proton beam to make the proton beam less diffusive, i.e. less
energy lost and scattering. However, the opened window can make the
plasma squirt outwardly. We use the foam with density $62\mathrm{mg/cc}$,
and the foam is divided by a layer with slanted or non-slanted slots,
as shown in Fig. \ref{fig:setup-layer1} and \ref{fig:setup-layer2}.
The end cap is $1\mathrm{g/cc}$ plastic. The shape of the slots,
the material and the thickness of the layer, and the material of the
wall are listed in Table \ref{tab:parameters-runs}. Some targets
are built with a pepper-pot screen (PPS)\citep{PPS_Brunetti2010},
as shown in Fig. \ref{fig:setup-pps}. The PPS is used for a narrow
view of the proton deflection signal in proton radiography, reducing
the signal contamination from off-center line-of-sight. The $200\mathrm{\mu m}$
diameter hole in the middle allows proton beams to go through the
central part of the target. Other holes are used as references to
register the position of protons. The PPS is a 40 $\mathrm{\mu m}$
thick tantalum foil.

\begin{table*}
\caption{The parameters and the maximum values of magnetic field and electron
temperature for the three different targets/runs we use. $T_{e}$
and $B$ are calculated by averaging over a $(200\mathrm{\mu m)^{2}}$
around the center of the target in the $x-z$ plane. PPS stands for
pepper-pot screen.\label{tab:parameters-runs}}

\begin{tabular}{|c|c|c|c|c|c|c|c|}
\hline 
\makecell{Target/ \\Run label} & \makecell{slanted\\slots} & \makecell{layer\\thickness} & \makecell{layer\\material} & \makecell{wall\\thickness} & \makecell{wall\\material} & \makecell{$T_e$(eV) \\ at 10ns} & \makecell{$B$(kGauss) \\ at 10ns}\tabularnewline
\hline 
\hline 
{\small{}A} & {\small{}Yes} & {\small{}$15\mathrm{\mu m}$} & {\small{}Mg} & $100\mathrm{\mu m}$ & {\small{}Be} & 25 & 158\tabularnewline
\hline 
{\small{}B} & {\small{}No} & {\small{}$6\mathrm{\mu m}$} & {\small{}Cu} & $150\mathrm{\mu m}$ & {\small{}CH} & 26 & 152\tabularnewline
\hline 
{\small{}C} & {\small{}No} & {\small{}$6\mathrm{\mu m}$} & {\small{}CH} & $150\mathrm{\mu m}$ & {\small{}CH} & 28 & 86\tabularnewline
\hline 
\end{tabular}
\end{table*}

In the initialization, the pressure of all the solid regions is $5\times10^{3}\mathrm{bar}$($=5\times10^{9}\mathrm{erg/cm^{3}}$),
and the temperature is calculated self-consistently from the equation
of state table. Using the same pressure instead of the same temperature
among all the solid regions can prevent one solid region from expanding
into another solid region and launching artificial shocks before the
high-energy-density conditions is reached. Under HED condition, the
pressure is larger than $10^{5}\mathrm{bar}$($=10^{11}\mathrm{erg/cm^{3}}$),
thus the initial pressure is low enough to have neglectable effect
on the simulations. The vacuum region is initially filled with $10^{-6}\mathrm{g/cc}$
helium to avoid numerical problems in hydrodynamics or MHD solvers.
The density is low enough that the effect of helium on the simulations
is negligible.

A 3D cartesian grid with $(240\times240\times464)$ zones is used
to resolve a $(1440\mathrm{\mu m}\times1440\mathrm{\mu m}\times2784\mathrm{\mu m})$
domain, corresponding to $6\mathrm{\mu m}$ per cell width. Using
AMR, each zone is adaptively refined to one leaf level, i.e. a resolution
of $3\mathrm{\mu m}$ or $2^{3}=8$ zones, if the mass fraction of
the layer material is larger than 10\%. The refinement allows us to
efficiently resolve the dynamics near the layer and reduce the computing
time spent on the zones far away from the layer. Although we cannot
resolve the turbulence dissipation scale with the current computing
capability and neither do we use Reynolds-averaging Navier-Stokes
(RANS) models such as BHR model to resolve the small scale dissipation
process of the fluid, FLASH is still a suitable tool for designing
these experiments because the fabricated layers have low surface roughness.

To model the laser driven energy deposition, we use the spatial and
temporal specifications of each of the 16 OMEGA driver beams. Ray
tracing by solving the geometric optics and the inverse bremsstrahlung
absorption is used. The 16 driver beams are turned on and turned off
simultaneously with a 1ns pulse duration and 8 beams on each side
of the target. Each delivers 500J of energy on a target. The radius
of each beam is 283$\mathrm{\mu m}$ and the intensity distribution
we use is gaussian.

For convention, $t=0$ is the time for laser turn on. The axis of
the shock tube is the $z$ axis. The layer dividing the foam is in
the $y-z$ plane, i.e. the plane with $x=0$ everywhere. The center
of the target is at $x=y=z=0$. The $x$ axis extends through the
window.

The primary diagnostic for temporally and spatially resolved profile
of the density and shock propagation in the experiments is the point
projection X-ray radiography with a vanadium backlighter at 23$\times$
magnification. The backlighter source emits 5180eV and 5205eV helium
like lines\citep{XRAY_thompson2001x}. The images are recorded on
the X-ray framing camera (XRFC)\citep{XRFC_Bradley1992,XRFC_Bradley1995,XRFC_Benedetti2012}.
We use SPECT3D\citep{SPECT3D_MacFarlane2007}\footnote{Prism Computational Sciences SPECT3D Overview \href{http://www.prism-cs.com/Software/Spect3D/overview.html}{http://www.prism-cs.com/Software/Spect3D/overview.html}}
to generate the synthetic ray-tracing X-ray image. The line of sight
of XRFC is along the $y$ axis, which captures the distortion of the
layer.

Proton radiography\citep{Prad_Li2006}, using $\mathrm{D^{3}He}$
(14.7 MeV) protons from fusion, measures magnetic fields. The temporal
resolution of proton radiograph is typically $\sim150\mathrm{ps}$
and the spatial resolution is typically $\sim45\mathrm{\mu m}$. The
diffusion of the proton beam caused by Coulomb scattering\citep{CScat_Moliere1948,CScat_Bethe1953}
and stopping power\citep{Stopping_Bethe1930,Stopping_strag_Bonderup1971,Stopping_Li1993,Stopping_GERICKE2002,Stopping_WDM_Zylstra2015}
is significant for the targets we use. We use Monte Carlo code MPRAD\citep{MPRAD_Lu2019}
to model the synthetic proton radiography, including the Lorentz force
and the effects from Coulomb scattering and stopping power. The proton
source stands at $(-0.75\mathrm{cm},0,0)$, while the image plate
CR39 is located 27cm from the center on the other side. The line of
sight of the proton radiography is perpendicular to the line of sight
of the X-ray image. The energy distribution of the proton source we
use in the simulation is a gaussian distribution with $\mathrm{FWHM}=0.25\mathrm{MeV}$
centered at $14.7\mathrm{MeV}$.

\section{Simulation results\label{sec:Simulation-results}}

\begin{figure*}
\includegraphics[scale=0.54]{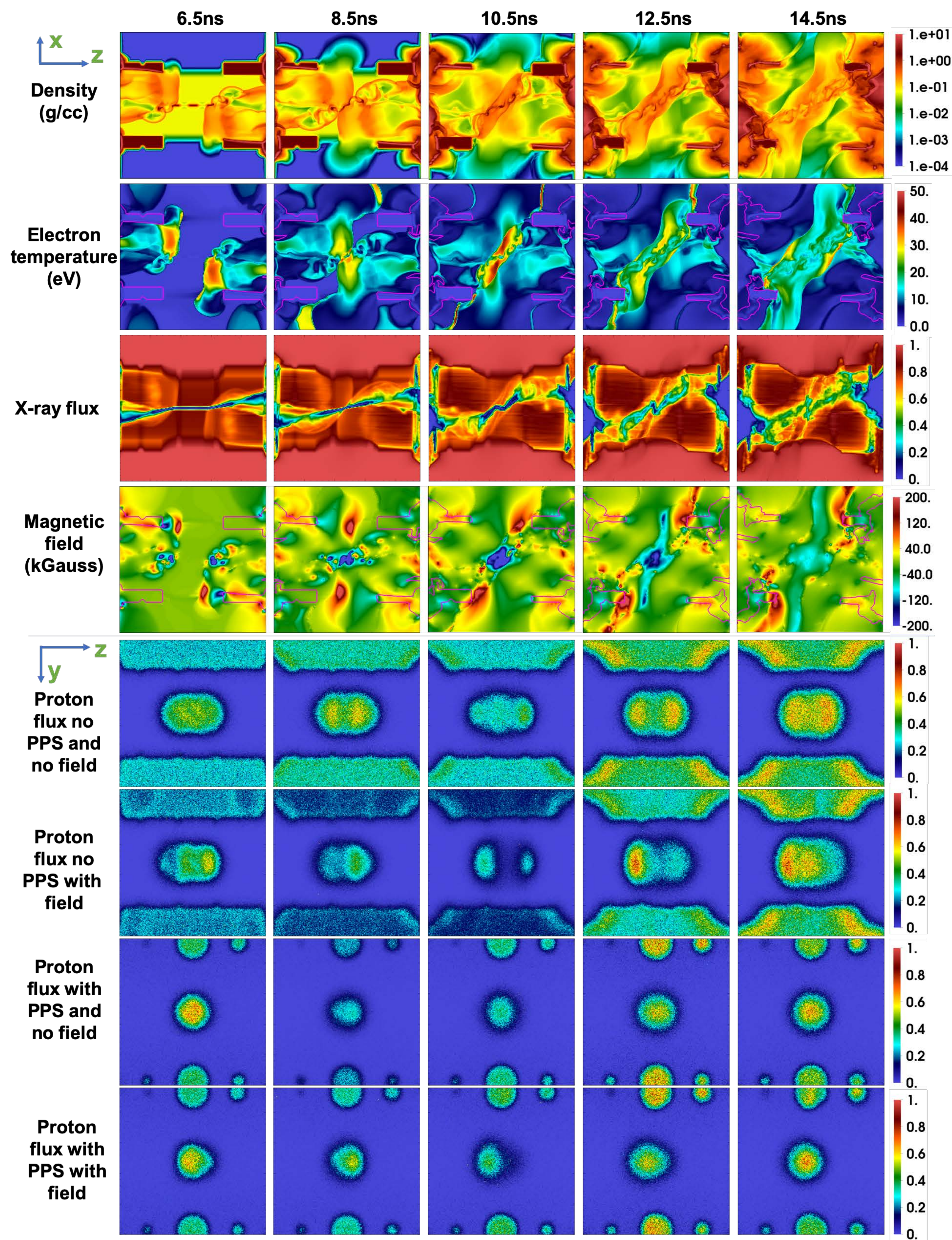}

\caption{Spatial distribution of different quantities at different time. The
size of all plots is $1200\mathrm{\mu m}\times1200\mathrm{\mu m}$.
From first to fourth row are: density at the $y=0$ plane, electron
temperature at the $y=0$ plane, X-ray flux normalized by the purely
transparent flux, magnetic field $B_{y}$ in kGauss at $y=0$ plane
(positive for into the plane). The plots in the second and the fourth
rows are overlaid with magenta contours for the density of the wall
material equal to $0.5\mathrm{g/cc}$. From fifth to the last rows
are proton images for four different cases as labeled. \label{fig:runA}}
\end{figure*}

\begin{figure*}
\includegraphics[scale=0.54]{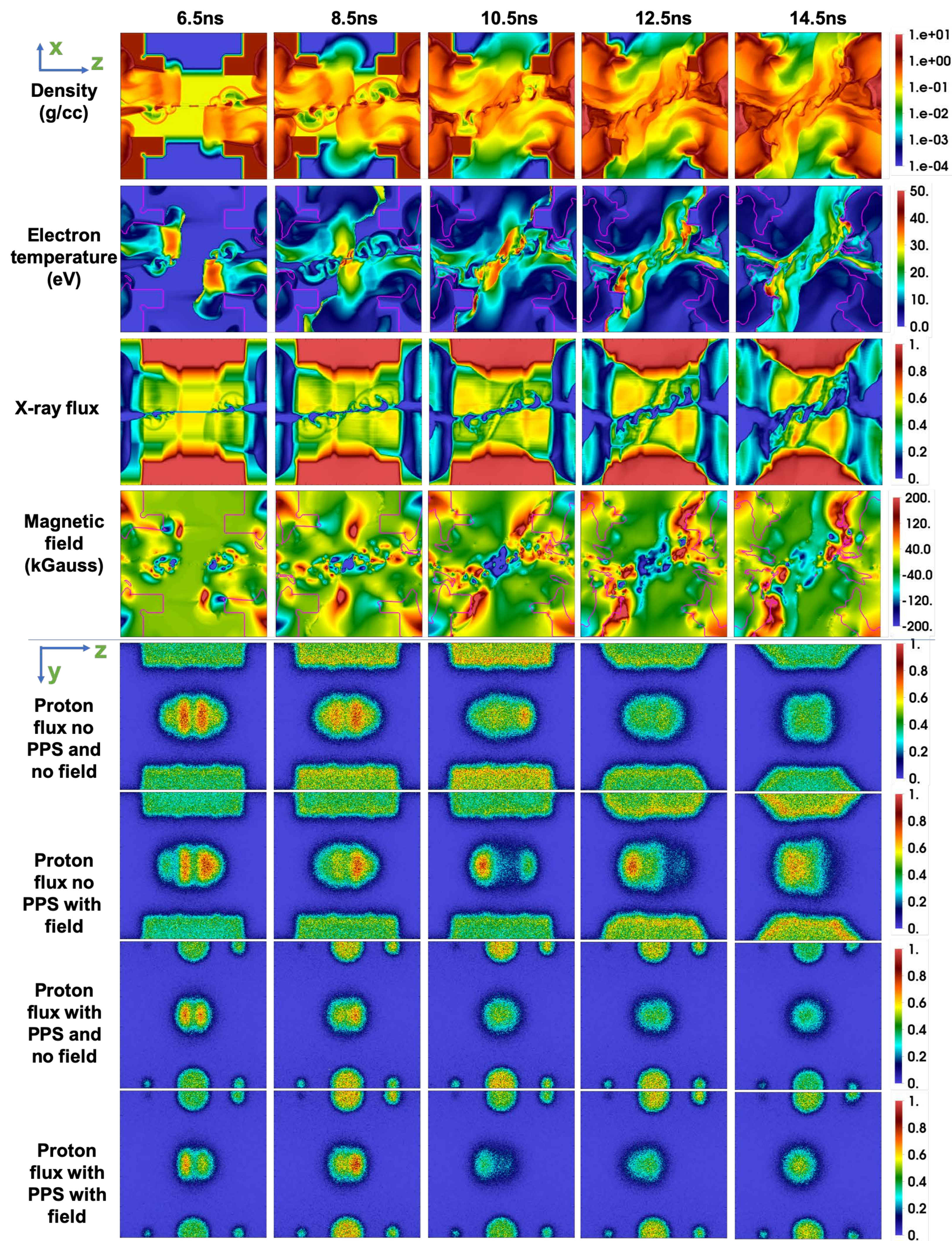}

\caption{Same as Fig. \ref{fig:runA} but for runB\label{fig:runB}}
\end{figure*}

\begin{figure*}
\includegraphics[scale=0.54]{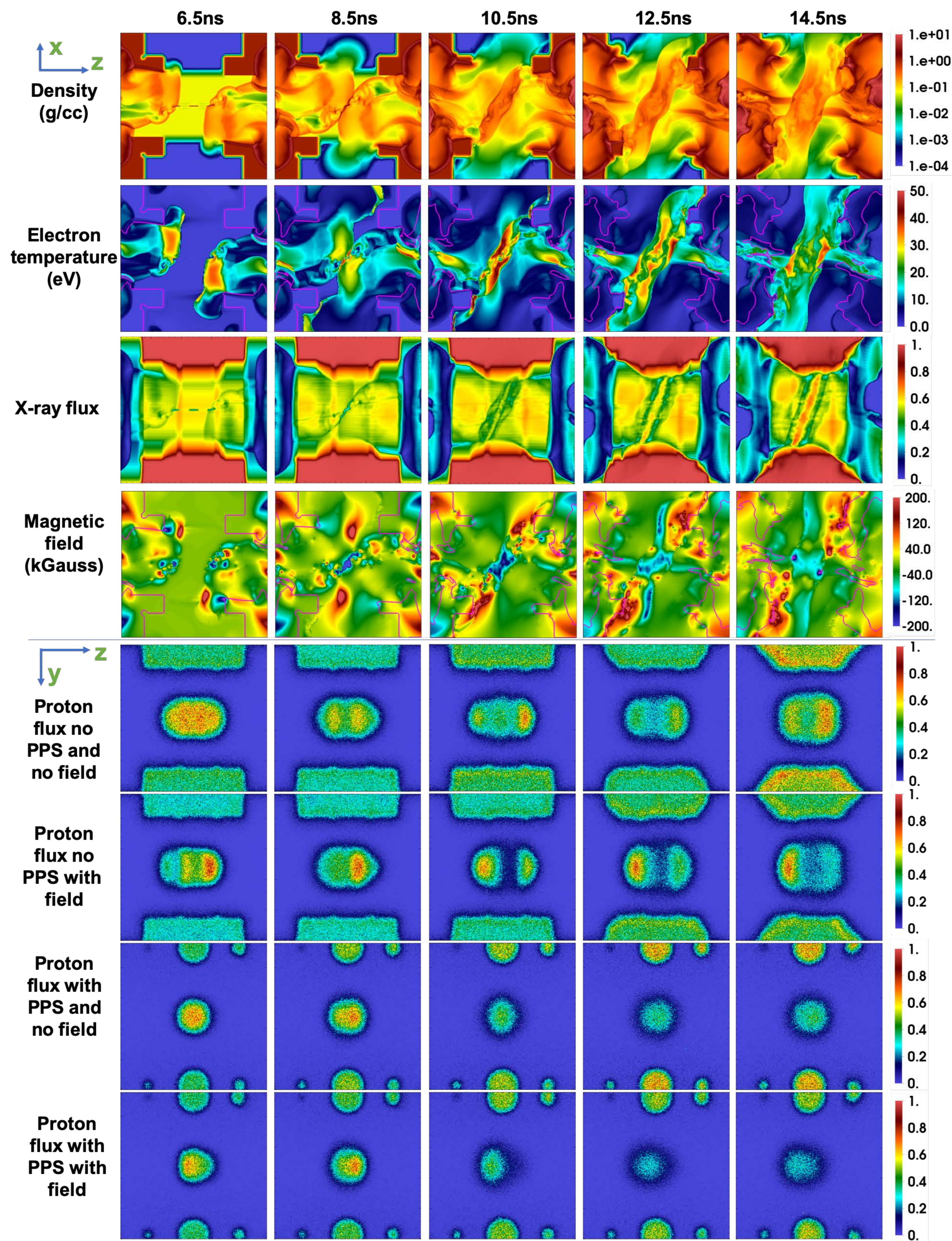}

\caption{Same as Fig. \ref{fig:runA} but for runC\label{fig:runC}}
\end{figure*}

We show the results from FLASH simulations and the synthetic radiography
to study the evolution and dynamics of the flows in the shock-shear
targets in Fig. \ref{fig:runA} to \ref{fig:runC}. In the synthetic
radiographs, the spatial scales of the synthetic radiographs are divided
by the magnification to align with the scales on the target system.
The target we use in this work are different from previous shock-shear
experiments\citep{SS_Doss2013a,SS_Doss2013b,SS_Merritt2017,SS_Merritt2015}
mainly in two aspects: (1) cut slots in the layer for seeding density
perturbation, (2) opened window on the wall for reducing the diffusion
of proton beams.

\begin{table*}
\caption{Simulated plasma properties for runA. All quantities are in cgs units
except temperature, which is expressed in eV. The length scale, $L$
is approximately the diameter of the tube ($\approx500\mathrm{\mu m}$).
The $n_{e}$, $\rho$, $T_{e}$ and $T_{i}$ are calculated by averaging
over a $(200^{2}\mathrm{\mu m})^{2}$ square around the center of
the target in the $x-z$ plane, at $t=10\mathrm{ns}$. The flow speed
is $u=7\times10^{6}\mathrm{cm/s}$ for each counter propagating flow.
\label{tab:Simulated-plasma-properties}}

\begin{tabular}{|c|c|c|}
\hline 
Plasma property & Formula & Value at $r=0$\tabularnewline
\hline 
\hline 
Electron density $n_{e}$($\mathrm{cm}^{-3}$) & $\cdots$ & $5.6\times10^{22}$\tabularnewline
\hline 
Mass density $\rho$($\mathrm{g/cm^{3}}$) & $\cdots$ & $0.36$\tabularnewline
\hline 
Electron temperature $T_{\mathrm{e}}$(eV) & $\cdots$ & $25$\tabularnewline
\hline 
Ion temperature $T_{\mathrm{i}}$(eV) & $\cdots$ & $25$\tabularnewline
\hline 
Magnetic field $B$ (gauss) & $\cdots$ & $1.6\times10^{5}$\tabularnewline
\hline 
Average ionization $Z$ & $\cdots$ & $1.9$\tabularnewline
\hline 
Average atomic weight $A$ & $\cdots$ & $7.3$\tabularnewline
\hline 
Flow speed $u$(cm/s) & $\cdots$ & $7\times10^{6}$\tabularnewline
\hline 
Sound speed $c_{s}$(cm/s) & $\frac{9.8\times10^{5}[ZT_{\mathrm{e}}+1.67T_{\mathrm{i}}]^{1/2}}{A^{1/2}}$ & $3.4\times10^{6}$\tabularnewline
\hline 
Mach number $M$ & $u/c_{s}$ & $2$\tabularnewline
\hline 
Coulomb logarithm $\ln\Lambda$ & $\min(23.5+\ln(T_{e}^{1.5}/n_{e}^{0.5}/Z),25.3+\ln(T_{e}/n_{e}^{0.5}))$ & $1.4$\tabularnewline
\hline 
Hall parameter $\chi$ & $6.1\times10^{12}\frac{T_{e}^{3/2}B}{Zn_{e}\ln\Lambda}$ & $8\times10^{-4}$\tabularnewline
\hline 
Plasma $\beta$ & $\frac{2.4\times10^{-12}n_{e}(T_{e}+T_{i}/Z)}{B^{2}/(8\pi)}$ & $5\times10^{3}$\tabularnewline
\hline 
Péclet number $Pe$ & $uL/(\frac{\kappa_{e}}{\frac{3}{2}n_{e}k_{B}})(\frac{\kappa_{e}}{\frac{3}{2}n_{e}k_{B}}=5.5\times10^{21}\frac{T_{e}^{5/2}}{n_{e}(3.3+Z)\ln\Lambda})$ & $8.3\times10^{3}$\tabularnewline
\hline 
Magnetic Reynolds number $Rm$ & $uL/\eta$$\bigg(\eta=8.2\times10^{5}\frac{(0.33Z+0.18)\ln\Lambda}{T_{e}^{3/2}}\bigg)$ & $47$\tabularnewline
\hline 
Reynolds number $Re$ & $uL/\nu$$\bigg(\nu=1.9\times10^{19}\frac{T_{\mathrm{i}}^{5/2}}{A^{1/2}Z^{3}n_{e}\ln\Lambda}\bigg)$ & $8.6\times10^{6}$\tabularnewline
\hline 
\end{tabular}
\end{table*}

\begin{figure}
\includegraphics[scale=0.5]{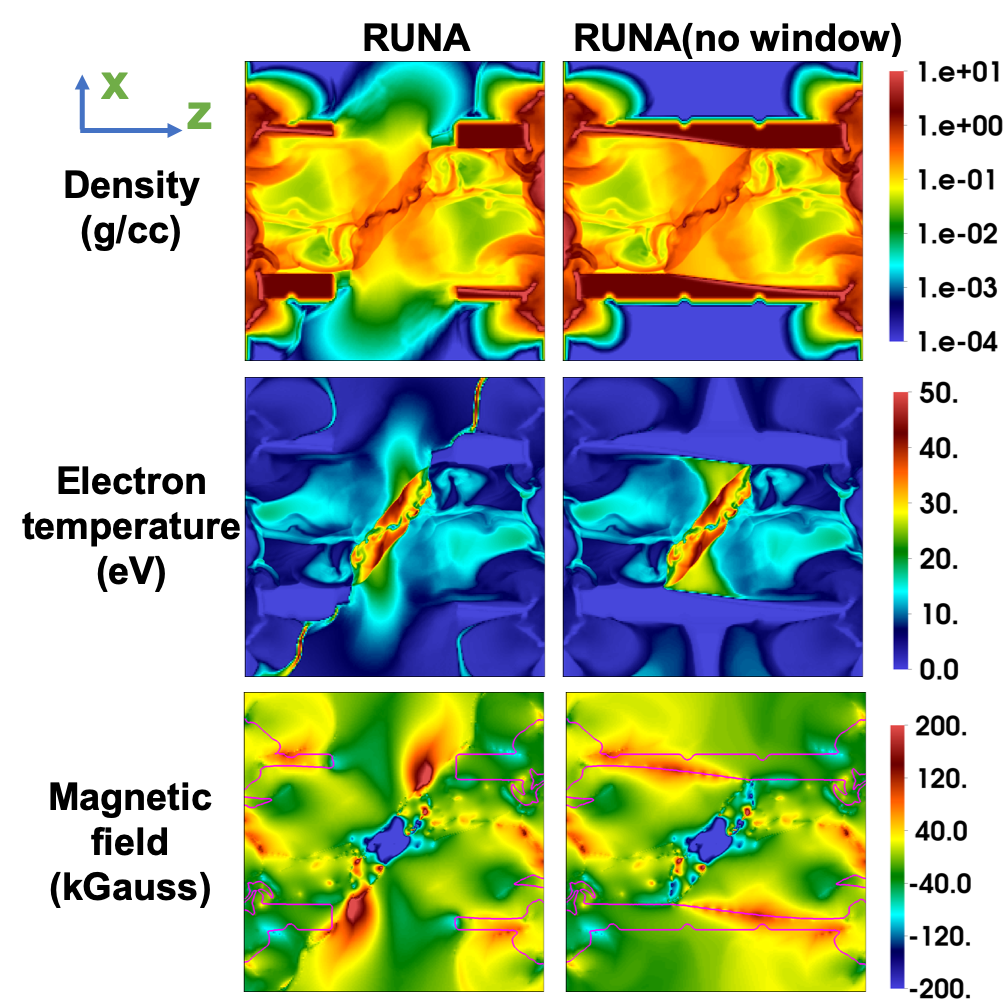}

\caption{Spatial distribution of different quantities for runA with or without
window at 10$\mathrm{ns}$. The size of all plots is $1200\mathrm{\mu m}\times1200\mathrm{\mu m}$.
From first to last row are: density at the $y=0$ plane, electron
temperature at the $y=0$ plane, magnetic field $B_{y}$ in kGauss
at $y=0$ plane (positive for into the plane). The plots in the second
row are overlaid with magenta contours for the density of the wall
material equal to $0.5\mathrm{g/cc}$.\label{fig:runAn}}
\end{figure}

\begin{figure}
\includegraphics[scale=0.6]{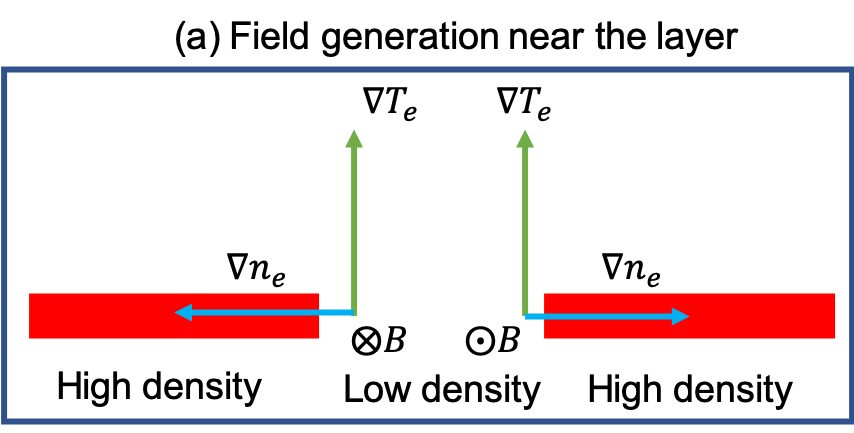}

\includegraphics[scale=0.6]{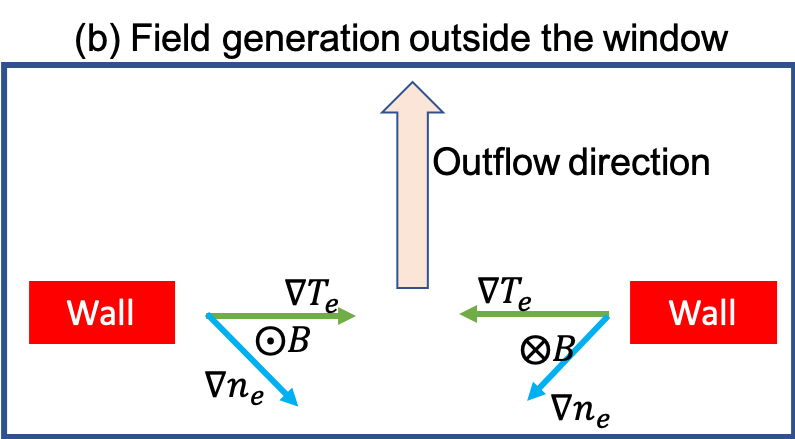}

\caption{Schematics of the magnetic field generation by Biermann battery term($\nabla n_{e}\times\nabla T_{e}$).
(a) Near the layer, the temperature gradient is perpendicular to the
layer due to thermal conduction, the density gradient is alternating
and along the layer due to the cut slots on the layer, so that the
Biermann generated field is alternating into and out of the plane.
(b) Outside the window, the density gradient points to the dense part
of the plume, the temperature gradient along the outflow direction
is small due to conduction, but the temperature gradient perpendicular
to the outflow direction survives due to continuous launching of the
plume from the shock tube, thus the field is into the plane on the
right side and out of the plane on the left side. \label{fig:Biermann}}

\end{figure}

\begin{figure}
\includegraphics[scale=0.7]{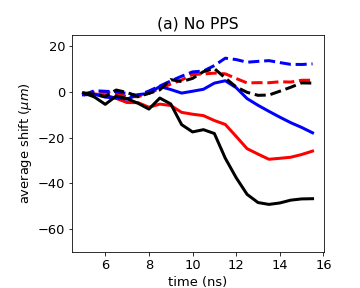}

\includegraphics[scale=0.7]{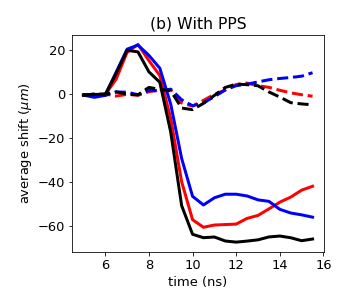}

\caption{The evolution of the averaged position of protons in the blob in final
energy range $14.3\mathrm{MeV}$ to $14.5\mathrm{MeV}$. The scale
is divided by the magnification to align with the scales on the target
system. The red curves are for runA, the black curves are for runB,
and the blue curves are for runC. The dashed curves are for the MPRAD
runs with magnetic field turned off, and the solid curves are for
MPRAD runs with magnetic field turned on. (a) is for no PPS case and
(b) is for with PPS case. \label{fig:shift}}
\end{figure}

\subsection{Hydrodynamics}

We show the evolution of density, electron temperature and X-ray flux
in the first three rows in Fig. \ref{fig:runA} to \ref{fig:runC}.
The gold plugs hold back the shock at one end of each half-cylinder
of foam. Two shocks of roughly same strength in the same material
propagate from opposite directions towards the center of the tube.
The layer placed in the middle between the two regions collimates
the shocked flows and introduces a length scale through its thickness
which will influence the dominant modes of the resulting shear instability.
The cut slots in the layer introduce alternating density gradients
and causes magnetic field generation by Biermann battery term, which
is discussed in Sec. \ref{subsec:Magnetic-fields}. Because the layer
does not fully collimate the shocks, oblique shocks are launched into
the opposite volumes of the tube. The shock front near the end of
the tube travels further transversely. It takes roughly $8.5\mathrm{ns}$
for the shocks to cross and create the pressure-balanced shear mixing
region. The pressure in the two regions is roughly equal and the shocked
material is the same on each side of the mixing layer, so that the
mixing region does not experience a net translation away from the
center of the shock tube. After $8.5\mathrm{ns}$, the oblique shock
on either end of the tube gradually crosses the primary shock from
the other direction. An oblique region of high density is developed
by the reverse shock.

The ideally constructed target should be symmetric about a rotation
of 180 degrees. However, the different effective laser intensities
on two ends of the target due to different laser incident angles cause
the two shocks to move at slightly different speeds. The shock from
the right side in Fig. \ref{fig:runA} to \ref{fig:runC}, moves slightly
faster. This asymmetry does not affect the overall picture of the
hydrodynamical and magnetic field evolution, but the asymmetry of
the density distribution can affect the proton radiography which is
discussed in Sec. \ref{subsec:Proton-radiography}.

Because of the opened window on the wall, there are plasma plumes
traveling outside the window. As shown in Fig. \ref{fig:runAn}, the
overall picture of hydrodynamical evolution is still similar to previous
shock-shear experiments without a window\citep{SS_Doss2013a,SS_Doss2013b,SS_Merritt2017,SS_Merritt2015},
although the plasma plume carries mass and energy away from the tube.
At later times, the shock can penetrate through the wall. This results
in plumes outside the wall, which can then interact with the plume
from the window.

\subsection{X-ray images}

The transmitted X-ray flux is shown in the third rows in Fig. \ref{fig:runA}
to \ref{fig:runC}. In the X-ray flux, the location and the shape
of the shock front is consistent with the density distribution and
can be easily identified. The shocks in the wall can also be seen
in the X-ray image. The plume launched from the wall or the window
has low density and is not visible in the X-ray flux. The layer has
high density and low X-ray transmission, leading to the low flux on
the X-ray image. For runA and runB, where the layer material is magnesium
and copper respectively, the contrast of X-ray flux between the layer
and the foam is high, while for runC where the layer material is CH,
the X-ray contrast is low.

\subsection{Magnetic fields\label{subsec:Magnetic-fields}}

When the shock from one end of the tube passes, the temperature is
high near the center of the half-cylinder as shown in the second rows
in Fig. \ref{fig:runA} to \ref{fig:runC}. A cold region is left
behind the shock. The temperature gradient near the layer is perpendicular
to the layer and pointing towards the shocked region, due to electron
heat conduction. The density gradient is alternating, caused by the
cut slots on the layer. Thus the Biermann battery term generates the
alternating magnetic field in the $\pm y$ direction, as shown in
Fig. \ref{fig:Biermann}(a). However, the cold region left behind
the shock has low electron temperature and thus high resistivity.
The magnetic fields behind the shock diffuse very quickly. In the
end, the only significant field left near the center of the tube is
in the $-y$ direction, because near the center of the tube, the layer
is at high density instead of at a cut slot. On both sides of that
high density layer, the field generation is in the $-y$ direction.
Two shocks from two ends of the tube cross, amplify the magnetic field
and create a doubly shocked, high temperature region, which has low
resistivity and the field is less diffusive.

The magnetic field in the plume traveling outside the window is generated
in a similar way to the magnetic field generated in the ablation plume
of a laser interaction with a solid target\citep{Prad_Li2006,PRAD_BB_Cecchetti2009,PRAD_BB_Li2007,PRAD_BB_Li2009}.
The plume is continuously launched by the flow inside the shock tube
and expands in all directions, with the density gradient to point
towards the dense part of the plume, as shown in Fig. \ref{fig:Biermann}(b).
The temperature gradient along the outflow direction is reduced due
to electron thermal conduction, but the temperature gradient perpendicular
to the outflow direction survives due to continuous launching of the
plume from the shock tube. Thus the magnetic field generated by the
Biermann battery term is into the plane on the right side and out
of the plane on the left side in Fig. \ref{fig:Biermann}(b).

The magnetic field evolution is shown in the fourth row in Fig. \ref{fig:runA}
to \ref{fig:runC}. In the center of the tube, a field pointing in
$-y$ direction dominates. Outside the window, the field pointing
in $+y$ direction survives, while the field pointing in $-y$ direction
diffuses quickly due to low temperature and high resistivity. The
total magnetic flux in the $y=0$ plane is conserved and vanishes.
We are interested in the magnetic field near the center of the tube
which can potentially affect the mix. The magnetic field outside the
window plays a role in the proton radiography as discussed in Sec.
\ref{subsec:Proton-radiography}, but we are not interested in its
dynamical importance because it is far away from the mix region. As
shown in Fig. \ref{fig:runAn}, the magnetic field near the center
of the tube is similar between the runs with and without the window.

\subsection{Proton radiography\label{subsec:Proton-radiography}}

We use the MPRAD code\citep{MPRAD_Lu2019} to simulate the proton
image by taking the output data from 3D FLASH simulations. In the
simulations, we use a typical size $45\mathrm{\mu m}$ for proton
source. We find that the features of the proton images are most prominent
in $14.3\mathrm{MeV}$ to $14.5\mathrm{MeV}$ band, i.e. protons losing
between $0.2\mathrm{MeV}$ and $0.4\mathrm{MeV}$ of kinetic energy.
We compare the proton images with/without field, and with/without
pepper pot screen (PPS) in the fifth to the last rows in Fig. \ref{fig:runA}
to \ref{fig:runC}. To quantify the asymmetry of the proton image,
the averaged horizontal proton position in the blob at the center
of the proton image is plotted in Fig. \ref{fig:shift}. The ideally
constructed target should be symmetric about a rotation of 180 degrees
and the proton image should also be symmetric in the absence of magnetic
field. The asymmetry of the proton image about the vertical axis can
be interpret as the existence of magnetic field.

However, in the no PPS case, i.e. the fifth rows in Fig. \ref{fig:runA}
to \ref{fig:runC}, the blob in the middle of the image can be slightly
asymmetric even without magnetic field. This asymmetry is not as large
as the asymmetry in the images where there is field but no PPS, i.e.
the six rows, which means the proton deflection by magnetic field
causes more asymmetry than by the density asymmetry due to the fact
that the shock from the right side in Fig. \ref{fig:runA} to \ref{fig:runC},
moves slightly faster. This slight difference is caused by the different
effective laser intensities on two ends of the target due to different
laser incident angles. In the simulations in this work, we do not
take into account the unevenness of the foam and the power imbalance
on two ends of the tube, which can potentially cause more the asymmetry
on the proton image than what we show in this work.

One advantage of using PPS is that the viewing of the surrounding
holes is through the regions without the field and the viewing of
the hole in the middle is only thorough the region with magnetic field,
so that the net deflection caused by the magnetic field can be determined
without another control shot using same target. With PPS, the asymmetry
in the no field case, i.e. the seventh row in Fig. \ref{fig:runA}
to \ref{fig:runC}, is significantly less than the without field and
without PPS case, i.e. the fifth rows. The PPS is very efficient in
reducing the asymmetry of the proton image cause by the intensity
imbalance on two ends and the unevenness of the foam. As shown in
Fig. \ref{fig:shift}(b), the asymmetry caused by the proton deflection
is significantly larger than that caused by the ununiform density.
The blob has a positive net shift at early time, because of the field
pointing in $+y$ direction in the plume outside the window. At about
$8.5\mathrm{ns}$, the proton deflection caused by the field pointing
in $+y$ direction in the plume outside the window and by the field
in near the center of the tube pointing in $-y$ direction cancels,
resulting in zero net shift of the blob on the proton image. At a
late time $t>10\mathrm{ns}$, the field pointing in $+y$ direction
moves away from the $z=0$ plane, but the field near the center of
the tube has no net advection, and the net shift of the blob is negative.
The shift value on the image plate divided by the magnification can
reach $50$ to $70\mathrm{\mu m}$. The difference between the early
time shift and late time shift can reach $70$ to $90\mathrm{\mu m}$.
The prediction for the net shift of the blob will be compared to the
experimental data to validate the magnetic field model in FLASH.

\section{Conclusions and discussions\label{sec:Conclusions-and-discussions}}

We carried out the radiation-MHD simulations and predicted the X-ray
and proton images by synthetic radiographs. The hydrodynamical evolution
can be measured using XRFC and compared with the simulation results.
The predicted proton radiography shows the direction and the amount
of the shift of the proton beam going through the window and/or PPS.
Although the target can diffuse the proton beam significantly, the
evolution of the shift in the synthetic proton radiography is still
consistent with the evolution of the magnetic fields in the target
system and shows change between early time and late time. However,
the prediction only shows the signal contribution from the mean magnetic
fields from different columns along the line of sight. The signal
from small scale fields always gets damped by the diffusion of the
proton beam. High energy proton beam accelerated by Target Normal
Sheath Acceleration (TNSA) mechanism using OMEGA EP beam experiences
less diffusion through the target\citep{Prad_Zylstra2012}. The Coulomb
scattering angle is roughly proportional to $E_{p}^{-2}$ where $E_{p}$
is the kinetic energy of the proton\citep{CScat_Moliere1948,CScat_Bethe1953,MPRAD_Lu2019}.

The simulation shows that the design we use can achieve a regime with
high plasma beta $\beta$. The Hall parameter $\chi$, defined by
the radio of electron gyro-frequency to electron collision frequency,
is small. The Reynold number $Re$ is high enough to ensure turbulence,
and the magnetic Reynolds number $Rm$ is around $50$. Under the
condition with these dimensionless parameters, the magnetic field
remains dynamically unimportant. The magnetic energy density from
Table \ref{tab:Simulated-plasma-properties} is $10^{9}\mathrm{erg/s}$,
which is only $0.3\%$ of the turbulent kinetic energy reported in
the simulation in Ref. \citep{SS_Doss2013b} for a previous mix modeling
for shock-shear targets under similar condition to this work. Thus
the magnetic field is also negligible for mix modeling in the shock-shear
targets. It is desirable to optimize the measurable magnetic fields
and improve the dynamical importance of the magnetic fields.

The Biermann battery generated magnetic field is roughly $\frac{ck_{B}T_{e}}{eLu}$
by balancing the Biermann battery term with the advection term. The
plasma beta $\beta$ is then proportional to $\frac{n_{e}T_{e}}{(T_{e}/Lu)^{2}}\propto\frac{n_{e}u^{2}}{L^{2}T_{e}}$.
If we keep the size of the target and the laser power, then $n_{e}u^{2}$
and $L$ are roughly constants, then $\beta\propto\frac{1}{T_{e}}$.
Thus increasing $T_{e}$ can reduce $\beta$ and make the Lorentz
force more important. The Hall parameter\citep{Braginskii1965} $\chi$
is proportional to $\frac{T_{e}^{3/2}}{n_{e}}$ and the magnetic Reynolds
number $Rm$ is proportional to $T_{e}^{3/2}$. Both $\chi$ and $Rm$
increase with temperature. For low $Rm$ and low magnetic Prandtl
number $Pr_{m}$, i.e. $Pr_{m}=Rm/Re\ll1$, the power spectrum of
the kinetic energy $E(k)$ and the power spectrum of the magnetic
energy $M(k)$ are related by $M(k)\propto k^{-2}E(k)$, and $M(k)$
is always softer than $E(k)$, and the magnetic field remains dynamically
unimportant even in small scales\citep{TMD_Schekochihin2007,TMD_Odier1998,TMD_Meinecke2014}.
High $Rm$ is favorable for the amplification of magnetic fields and
a hard power law for magnetic energy spectrum\citep{TMD_Schekochihin2007,TMD_Tzeferacos2017,TMD_Tzeferacos2018}.
One way to achieve a higher temperature is to lower the density of
the foam. However, making a low density foam in the target is challenging
for target fabrication. It causes the unevenness in the foam, leads
to the unevenness of the proton image, and makes it difficult to interpret
the experimental data from proton radiography. In a low density foam,
the flow may move too fast so that the time window for diagnostics
is narrow.

Some experiments\citep{RES_Shepherd1988} and theories\citep{RES_Renaudin2002,RES_Robinson2015}
show that around $10\mathrm{eV}$ the value of electrical resistivity
(electrical resistivity $\eta$ is related to magnetic resistivity
$\eta_{B}$ by $\eta_{B}=\frac{c^{2}}{4\pi}\eta$) is different from
the Spitzer resistivity. However, the electrical resistivity with
temperature and density dependency under the condition of our experiment
design is not well constrained. If the modeling in this work is correct
in terms of electrical resistivity, then this would indicate that
the magnetic field may not be dynamically important. However, if the
electrical resistivity is significantly lower than the Spitzer resistivity
that we use in this work, then the code underestimates the magnetic
fields, and the mix model could potentially cover up the magnetic
field effects by the choice of the initial input conditions for the
model. Future experiments executed at higher temperatures can potentially
make magnetic fields start to play a more important role. In the future
development of the simulations, the implementation of implicit method
for the magnetic diffusion equation is desirable for the case of large
resistivity where fully explicit method requires small time step.

\section{Acknowledgements}

Research presented in this paper was supported by the Laboratory Directed
Research and Development(LDRD) program 20180040DR of Los Alamos National
Laboratory(LANL). The simulations were performed with LANL Institutional
Computing which is supported by the U.S. Department of Energy National
Nuclear Security Administration under Contract No. 89233218CNA000001,
and with the Extreme Science and Engineering Discovery Environment
(XSEDE), which is supported by National Science Foundation(NSF) grant
number ACI-1548562.

\bibliographystyle{apsrev4-1}
\bibliography{HEDB}

\end{document}